\documentclass[pra,aps,showpacs,articles]{revtex4}
\usepackage{epsf,epsfig}
\usepackage[psamsfonts]{amssymb}
\usepackage{amsmath}

\newcommand{\ket}[1]{|#1\rangle}
\newcommand{\bra}[1]{\langle #1|}
\newcommand{\proj}[1]{\ket{#1}\bra{#1}}

\newcommand{\expect}[3]{\langle#1|#2|#3\rangle}
\newcommand{\outprod}[2]{\ket{#1}\bra{#2}}
\newcommand{\half}{\mbox{$\textstyle \frac{1}{2}$}}

\begin{document}

\title{Influence of Dephasing on the Entanglement Teleportation via
 a two-qubit Heisenberg XYZ system}
\author{S. Javad
Akhtarshenas \footnote{akhtarshenas@phys.ui.ac.ir}, Fardin
Kheirandish\footnote{fardin$_{-}$kh@phys.ui.ac.ir} and Hamidreza
Mohammadi \footnote{h.mohammadi@phys.ui.ac.ir}
}\affiliation{Department of Physics, University of Isfahan,
 Hezar Jarib Ave., Isfahan, Iran}

\begin{abstract}
The entanglement dynamics of an anisotropic two-qubit Heisenberg XYZ
system in the presence of intrinsic decoherence is studied. The
usefulness of such system for performance of the quantum
teleportation protocol ${\cal T}_0$ and entanglement teleportation
protocol ${\cal T}_1$ is also investigated. The results depend on
the initial conditions and the parameters of the system. For the
product and maximally entangled initial states, increasing the size
of spin-orbit interaction parameter $D$ amplifies the effects of
dephasing and hence decreases the asymptotic entanglement and
fidelity of teleportation. We show that the XY and XYZ Heisenberg
systems provide a minimal resource entanglement, required for
realizing efficient teleportation. Also, we find that for the some
special cases there are some maximally entangled states which are
immune to intrinsic decoherence. Therefore, it is possible to
perform the quantum teleportation protocol ${\cal T}_0$ and the
entanglement teleportation ${\cal T}_1$ with perfect quality by
choosing a proper set of parameters and employing one of these
maximally entangled robust states as initial state of the resource.
\end{abstract}

\maketitle



\section{INTRODUCTION}
Entanglement is a central theme in quantum information processing
as a uniquely quantum mechanical resource that plays a key role in
many of the most interesting applications of quantum computation
and quantum information \cite{NC-book,A-book, V-book}. Thus a
great deal of efforts have been devoted to study and characterize
entanglement in the recent years . The central task of quantum
information theory is to characterize and quantify entanglement of
a given system. A pure state  of pair of quantum systems is called
entangled if it is unfactorizable, e.g. singlet state of two
half-spin system. A mixed state $\rho$ of a bipartite system is
said to be separable or classically correlated if it can be
expressed as a convex combination of uncorrelated states  $\rho_A$
and $\rho_B$ of each subsystems i.e. $\rho=\sum_{i} \,\omega_i
\rho_A ^i \otimes \rho_B ^i$ such that $\omega_i \geq 0$ and
$\sum_{i}\,\omega_i =1$, otherwise $\rho $ is entangled \citep
{Werner,A-book}. Many measures of entanglement have been
introduced and analyzed \cite{NC-book,A-book, V-book}. Here we use
the negativity as a measure of entanglement. For the $\Bbb{C}^2
\otimes \Bbb{C}^2$ bipartite systems, the negativity of a state
$\rho$ is defined as
\begin{eqnarray} \label{negativity}
E_N (\rho)=2 \max\{-\lambda_{min},0\},
\end{eqnarray}
where $\lambda_{min}$ is the minimum eigenvalue of $\rho^{T_A}$,
and $T_A$ denotes the partial transpose with respect to the part A
of the bipartite system.

It is well known that a two-qubit entangled system can be used to
perform the quantum teleportation protocols \cite{BBCJPW}. The
pioneering authors of quantum infirmation theory have showed that
the mixed quantum channels which allow to transfer the quantum
information with fidelity larger than $\frac {2}{3}$ are
worthwhile \cite{Pop}. By using the isomorphism between quantum
channels and a class of bipartite states and twirling operations,
Horodecki \textit{et al.} have shown that the optimal fidelity of
teleportation for a bipartite state acting on $\Bbb{C}^2 \otimes
\Bbb{C}^2$ \cite{HHH} is
$\Phi_{max}=\frac{d{\cal{F}}_{max}+1}{d+1}$ where
${\cal{F}}_{max}$ is the fully entangled fraction of the resource.
Then, Bowen and Bose have shown that in the standard teleportation
protocol ${\cal T}_0$ with an arbitrary mixed state resource the
teleportation process can be considered as a general depolarizing
channel with the probabilities given by the maximally entangled
component of the resource and without additional twirling
operations \cite{BB}. Using the property of the linearity of
teleportation process, Lee and Kim \cite{BBCJPW} have shown that
quantum teleportation preserves the nature of quantum correlation
in the unknown entangled states if the channel is quantum
mechanically correlated. They have considered entanglement
teleportation of an entangled state via two independent, equally
entangled, noisy channel, represented by Werner states. In this
teleportation protocol ${\cal T}_1$, the joint measurement is
decomposable into two independent Bell measurements and the
unitary operation is also decomposable into two local Pauli
rotations. In other word, ${\cal {T}}_1$ is a straightforward
generalization of the standard teleportation protocol ${\cal
{T}}_0$ just doubling the setup. Lee and Kim found that the
quantum correlation of the input state is lost during the
teleportation even the channel has nonzero entanglement. They also
found that in order to achieve a favorite teleportation fidelity,
the quantum channel should possess a minimal entanglement. Hence,
in comparison with quantum teleportation, entanglement
teleportation imposes more stringent conditions on the quantum
channel \cite{LK}.

Unfortunately, decoherence destroys the quantumness of the system
and hence will decrease the useful entanglement between the parts
of the system \cite{M-book,BP-book, Hamid2}. There are several
approaches to consider the decoherence and solve the quantum to
classic transition problem. One of these approaches is based on
modifying the Schr\"odinger equation in such a way that the
quantum coherence is automatically destroyed as the system
evolves. This mechanism is called "intrinsic decoherence" and has
been studied in the framework of several models (see \cite{Moya}
and references therein). In particular, Milburn has proposed a
simple modification of the standard quantum mechanics based on the
assumption that for sufficiently short time steps the system
evolution is governed by a stochastic sequence of identical
unitary transformation rather than continuous unitary evolution
\cite{Milburn}. This assumption leads to a modification of the
Schr\"odinger equation which includes a term corresponding to the
decay of quantum coherence in the energy basis. Using a "Poisson
model" for stochastic time step, Milburn obtained the following
dynamical master equation in the first order approximation
\begin{eqnarray}\label{dynamics}
\frac{d}{dt} \rho(t)=-i [ H,\rho]-\frac{1}{2 \gamma}
[H,[H,\rho(t)]],
\end{eqnarray}
where $H$ is the Hamiltonian of the system, $\rho(t)$ indicates
the state of the system and $\gamma$ is the mean frequency of the
unitary step and determines the rate of decoherence
\cite{Milburn}. In the limit $\gamma \longrightarrow \infty$ the
Schr\"{o}dinger's equation is recovered. Note that, in this
mechanism of decoherence, the decay of quantum coherence is a
result of phase relaxation process, so in the following we will
only deal with dephasing processes without the usual energy
dissipation associated with normal decay. The first order
correction to the equation (\ref {dynamics}) leads to
diagonalization of the density operator in the energy eigenstate
basis,
\begin{eqnarray}\label{rho dot 2}
\frac{\partial}{\partial t}
\expect{\varepsilon'}{\rho(t)}{\varepsilon}
=-i(\varepsilon-\varepsilon')\expect{\varepsilon'}{\rho(t)}{\varepsilon}-
\frac{1}{2 \gamma}(\varepsilon-\varepsilon')^2
\expect{\varepsilon'}{\rho(t)}{\varepsilon}.
\end{eqnarray}
Note that the rate of diagonalization (dephasing) in the energy
basis depends on the square of the energy separation of the
superposed states. Thus the coherence between states that are
widely separated in energy, decays rapidly. A formal solution of
the Milburn's dynamical master equation (\ref{dynamics}) can be
expressed as \cite{Moya}
\begin{eqnarray} \label{formal solution}
\rho(t)=\sum_{k=0}^\infty M_k(t) \rho(0) M_k^\dagger(t),
\end{eqnarray}
where
\begin{eqnarray} \label{M-k}
M_k(t)=\sqrt{\frac{t^k}{k!}} H^k \exp(-i H t) \exp(-\frac{t}{2
\gamma} H^2).
\end{eqnarray}
It is evident from the master equation (\ref{dynamics}) that the
state of the system remains constant in time, if the initial state
of the system commutes with H. Thus all density matrices which can
be written as a classical mixture of the eigenstates
$\ket{\psi_i}$ of the Hamiltonian, i.e. $\rho=\sum\limits_i {p_i
\proj{\psi_i}}$ with $\sum\limits_i{p_i}=1$, are immune to
intrinsic decoherence. One important case is the thermal state of
the system with $p_i=\frac{e^{-\beta \varepsilon_i}}{tr[e^{-\beta
H}]}$, where $T=\frac{1}{ k_B \beta}$ is the temperature and $k_B$
is the Boltzman constant \cite{Hamid1}. In other words, the set of
all thermal states with different temperatures span an intrinsic
decoherence free subspace. A decoherence free subspace is a
Hilbert space such that each state of this space is immune to
decoherence \cite{M-book}.

The effects of intrinsic decoherence on the dynamics of
entanglement and quantum teleportation and entanglement
teleportation of Heisenberg systems have been studied in a number
of works \cite{Yeo,HXZ,GXS,GS}. For example Ye Yeo in Ref.
\cite{Yeo} has shown that in an anisotropic two-qubit XY
Heisenberg system, the nonzero thermal entanglement produced by
adjusting the external magnetic field beyond some critical
strength is a useful resource for teleportation via ${\cal T}_0$
and ${\cal T}_1$ protocols. The authors of Ref. \cite{HXZ} have
shown that, adjusting the magnetic field can  reduce the effects
of the intrinsic decoherence and accordingly one can obtains the
ideal fidelity of teleportation via XYZ Heisenberg systems. Also,
the results of Ref. \cite{GXS} showed that for an initial pure
state of the resource, which is the entangled state of a two-qubit
XXZ Heisenberg chain, an inhomogeneous magnetic field can reduce
the effects of intrinsic decoherence. Then the authors of Ref.
\cite{GS} argued that if the initial state is an unentangled
state, we can improve the fidelity of teleportation protocol
${\cal T}_0$ via two-qubit Heisenberg XXX systems in the absence
of the magnetic field by introducing the spin-orbit (SO)
interaction , arising from Dzyaloshinski-Moriya (DM) interaction.
However, the dynamics of entanglement and entanglement
teleportation of more involved spin systems has not been
discussed, yet.

In this paper, we investigate the influence of the intrinsic
decoherence (dephasing) on the entanglement dynamics and
teleportation scheme of a two qubit anisotropic Heisenberg XYZ
system under the influence of an inhomogeneous magnetic field and
in the presence of SO interaction. This system is suitable for
modelling of a system which is realized by the spin of two
electrons confined in two coupled quantum dots \cite{LDi, DD, CL}.
Because of weak vertical or lateral confinement, electrons can
tunnel from one dot to the other and spin-spin and spin-orbit
interactions between the two qubits exist. In summary, we show
that the dynamical and asymptotical behavior of the entanglement,
the quality of the quantum teleportation and the entanglement
teleportation and also the entanglement of the replica state, are
dependent on the initial conditions and the parameters of the
model. We discuss the problem for some special initial states and
investigate the role of the parameters of model (such as coupling
coefficients $J_\mu$, magnetic field $B$, inhomogeneity of
magnetic field $b$, SO interaction parameter $D$,...) on the
entanglement properties of the system. The results show that for
the product and maximally entangled initial states, the asymptotic
value of the entanglement decreases with $D$. The fidelity of the
teleportation approaches $2 \over 3$ form above for large values
of $D$, for both product and maximally entangled initial states of
the resource. Furthermore the results show that, it is possible to
perform the teleportation protocols ${\cal T}_0$ and ${\cal T}_1$
with perfect quality in the XY and XYZ Heisenberg systems if the
initial conditions and system parameters are set properly.

The paper is organized as follows. In Sec. II, we introduce the
Hamiltonian of a Heisenberg system under the influence of an
inhomogeneous magnetic field with taking into account the SO
interaction. For given initial states, the density matrix of the
system at a later time is derived exactly by solving the Milburn's
dynamical equation. The effects of initial conditions and system
parameters on the dynamics of entanglement, as measured by
negativity, is also studied in this section. The quantum
teleportation and entanglement teleportation processes via the
above system are investigated in the subsections II-A and II-B.
Finally, in Sec. III a discussion concludes the paper.

\section{THEORETICAL TREATMENT}
\begin{figure}
\epsfxsize=16cm \ \centerline{\hspace{0cm}\epsfbox{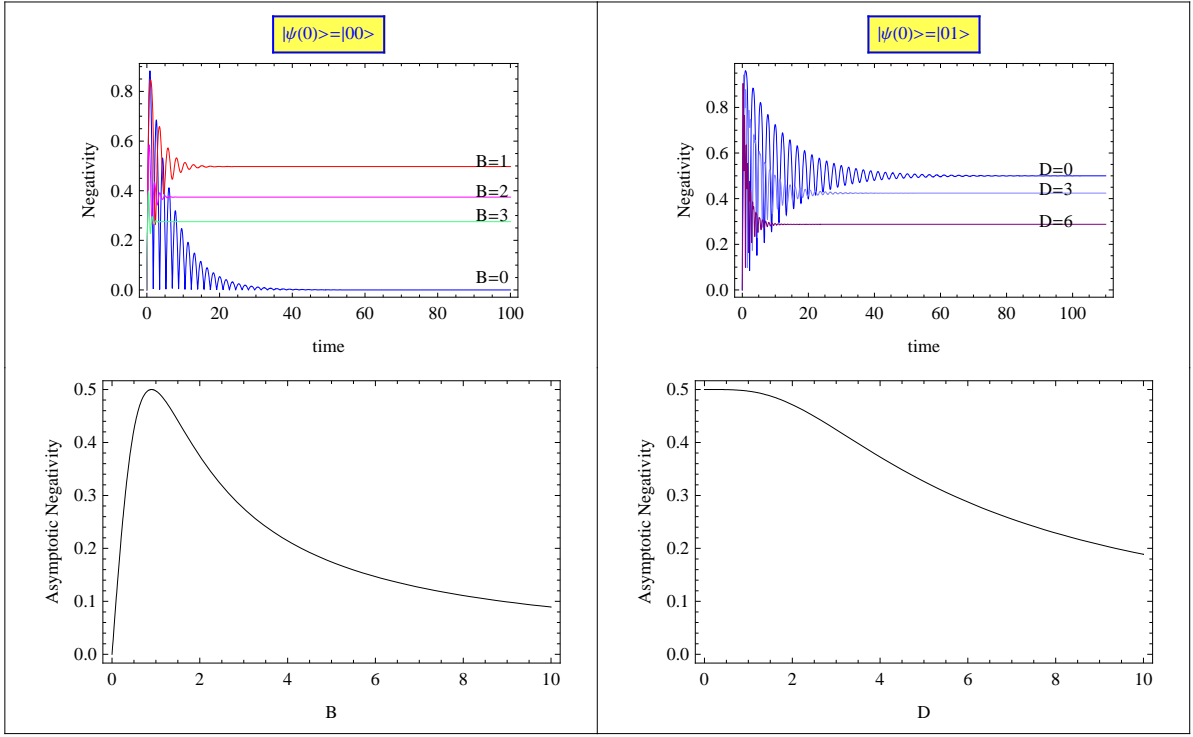}} \
\caption{(Color online) Dynamical and asymptotical behavior of the
negativity vs. the model parameters and for the product initial
states, $\ket{00}$ (left graph) and $\ket{01}$ (right graph). The
parameters of the model are chosen to be $J=1$, $\chi=0.9$,
$J_z=0.5$ and $b=1$. We have set $D=0$ and $\gamma=0.09$ for the
left and $B=3$ and $\gamma=0.02$ for the right graph. All parameters
are dimensionless.}\label{figure1}
\end{figure}

\begin{figure}
\epsfxsize=16cm \ \centerline{\hspace{0cm}\epsfbox{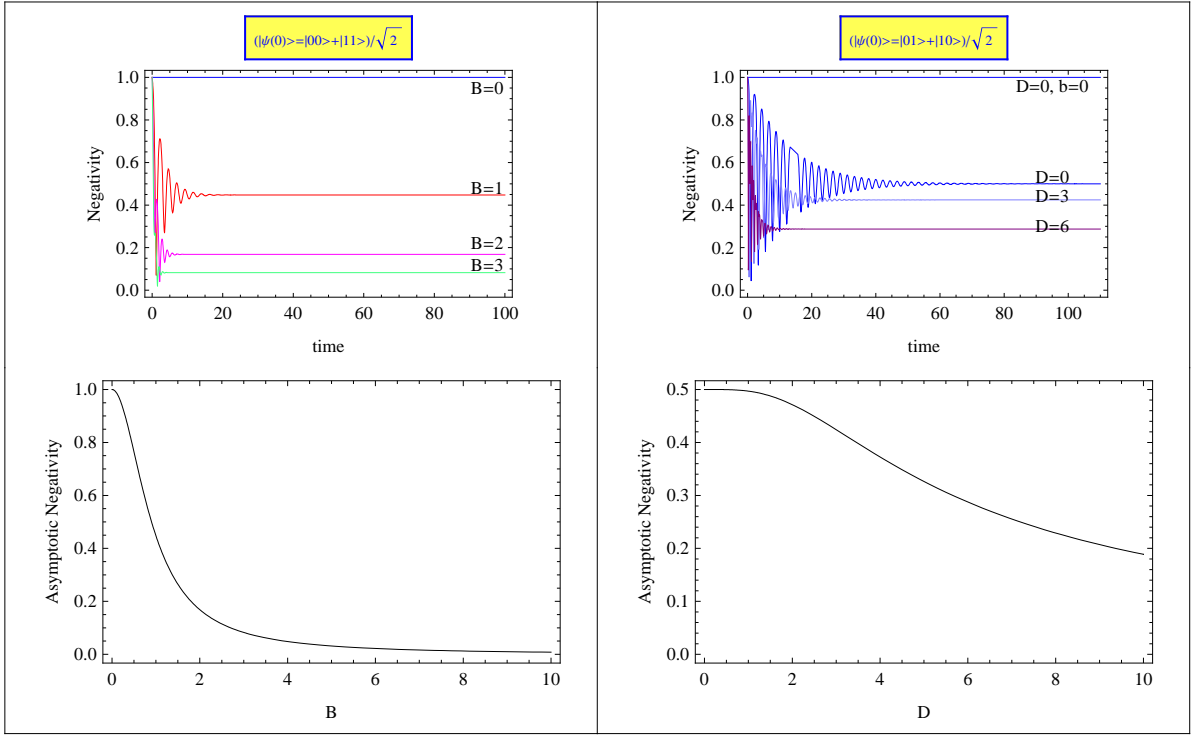}} \
\caption{(Color online) Dynamical and asymptotical behavior of the
negativity vs. the model parameters and for the maximally entangled
initial states,$\frac{\ket{00} \pm \ket{11}}{\sqrt 2}$ (left graph)
and $\frac{\ket{01} \pm \ket{10}}{\sqrt
 2}$ (right graph). The parameters are the same as in Fig. \ref{figure1}.
}\label{figure2}
\end{figure}

The Hamiltonian of a two-qubit anisotropic Heisenberg XYZ-model in
the presence of inhomogeneous magnetic field and spin-orbit
interaction is defined by \cite{Hamid1}
\begin{eqnarray}\label{Hamiltonian 1}
 H &=& {\textstyle{1 \over 2}}(J_x \,\sigma _1^x \sigma _2^x \,
 + J_y \,\sigma _1^y \sigma _2^y  + J_z \,\sigma _1^z \sigma _2^z
 +\textbf {B} _1 \cdot \boldsymbol {\sigma} _1
 + \textbf{B} _2 \cdot \boldsymbol {\sigma} _2
 \nonumber\\&+& \textbf{D} \cdot (\boldsymbol {\sigma} _1  \times
 \boldsymbol{\sigma}  _2
 )+ \delta \,\, \boldsymbol {\sigma}_1 \cdot \overline{\mathbf{\Gamma}}\cdot \boldsymbol
 {\sigma}_2),
 \end{eqnarray}
where $\boldsymbol{\sigma}_{j}=(\sigma^{x}_{j}, \sigma^{y}_{j},
\sigma^{z}_{j})$ is the vector of Pauli matrices, $\textbf{B}_j
\,(j=1,2)$  is the magnetic field on site j, $J_\mu \,(\mu=x,y,z)$
are the real coupling coefficients (the chain is
anti-ferromagnetic (AFM) for $J_\mu >0$ and ferromagnetic (FM) for
$J_\mu <0$)  and $\textbf{D}$ is  Dzyaloshinski-Moriya vector,
which is of first order in spin-orbit coupling and is proportional
to the coupling coefficients ($J_\mu$) and
$\overline{\mathbf{\Gamma}}$ is a symmetric tensor which is of
second order in spin-orbit coupling \cite{D,M1,M2,M3}. For
simplicity, we assume $\textbf{B}_j =B_j \, \boldsymbol{\hat{z}}$
such that $B_1 =B+b$ and $B_2 =B-b$, where b indicates the amount
of inhomogeneity of magnetic field. If we take $\textbf{D}= J_z D
\, \boldsymbol{\hat{z}}$ and ignore the second order spin-orbit
coupling, then the above Hamiltonian can be written as
\footnote{The parameters $D$ and $\delta$ are dimensionless. In
systems like coupled GaAs quantum dots $D$ is of order of a few
percent, while the order of the last term is $10^{-4}$ which is
negligible.}:
\begin{eqnarray}\label{Hamiltonian 2}
H &=& J\chi (\sigma _1^ +  \sigma _2^ +   + \sigma_1^ -  \sigma
_2^ -  ) + (J + iJ_z D)\sigma _1^ +\sigma _2^ -   + (J - iJ_z
D)\sigma _1^ -  \sigma _2^ + \nonumber\\&+& \frac{{J_z }}{2}\sigma
_1^z \sigma _2^z  + (\frac{{B + b}}{2})\sigma _1^z + (\frac{{B -
b}}{2})\sigma _2^z,
\end{eqnarray}
where $J :=\frac{J_x + J_y}{2}$, is the mean coupling coefficient
in the XY-plane, $\chi :=\frac{J_x - J_y}{J_x + J_y}$ specifies
the amount of anisotropy in the XY-plane (partial anisotropy,
$-1\leq \chi \leq 1$) and $\sigma^\pm=\frac{1}{2}(\sigma^x \pm
i\sigma^y)$ are lowering and raising operators. The spectrum of H
is easily obtained as
\begin{eqnarray}\label{spectrum}
\,H \ket{\psi ^ \pm}  = \varepsilon _{1,2} \ket{\psi ^ \pm }\,, \nonumber \\
\\ \nonumber H \ket{\Sigma^ \pm }  = \varepsilon _{3,4} \ket{\Sigma ^
\pm}\,,
\end{eqnarray}
where the eigenstates expressed in the standard basis
$\{\ket{00},\ket{01},\ket{10},\ket{11}\}$ are
\begin{eqnarray}\label{eigenstates}
\begin{array}{l}
\ket{\psi ^ \pm}  = N^ \pm  ( (\frac{{b \pm \xi }}{{J - iJ_z
D}})\ket{01}  + \ket{10})\,,
\\ \\
\ket{\Sigma ^ \pm}  = M^ \pm  ( (\frac{{B \pm \eta }}{{J\chi }})\ket{00}  + \ket{11} )\,, \\
\end{array}
\end{eqnarray}
with the eigenvalues
\begin{eqnarray}\label{eigenvalues}
\begin{array}{l}
\varepsilon _{1,2}  =  - \frac{{1}}{2} J_z \pm \xi\,,  \\ \\
\varepsilon _{3,4}  = \frac{{1}}{2} J_z \pm \eta\,,  \\
\end{array}
\end{eqnarray}
respectively. In the above equations $N^\pm
=\frac{1}{\sqrt{1+\frac{{(b \pm \xi)^2}}{J^2 +(J_z D)^2}}}$ and
$M^\pm =\frac{1}{\sqrt{1+(\frac{{B \pm \eta}}{J \chi})^2}}$ are
the normalization constants. Here we have defined, $ \xi  := \sqrt
{b^2  + J^2  + (J_z D)^2 }$ and
$\eta := \sqrt {B^2  + (J\chi )^2 }$, for later convenience.\\
According to Eq.(\ref{formal solution}) it is easy to show that,
the time evolution of the density operator $\rho(t)$ for the above
mentioned two qubit Heisenberg system which is initially in the
state $\rho(0)$, under intrinsic decoherence is given by
\begin{eqnarray}\label{rho t}
\rho(t)=\sum_{m,n=1}^{4}\exp[- \frac{\gamma t}{2}
(\varepsilon_m-\varepsilon_n)^2-i
(\varepsilon_m-\varepsilon_n)t]\,\expect{\phi_m}{\rho(0)}{\phi_n}\,\outprod{\phi_m}{\phi_n},
\end{eqnarray}
where eigenenergies $\varepsilon_{m,n}$ and the corresponding
eigenstates $\ket{\phi_{1,2}}=\ket{\psi^{\pm}}$ and
$\ket{\phi_{3,4}}=\ket{\Sigma^{\pm}}$ are given in
Eqs.(\ref{eigenstates}) and (\ref{eigenvalues}) and
$\gamma$ is the phase decoherence rate.\\
In the following we will examine the evolution of entanglement
under intrinsic decoherence of a class of bipartite density
matrices having the standard form
\begin{eqnarray}\label{initial dm}
\rho(0) &=& \mu_+ \proj{00}+ \mu_-\proj{11}+ \nu
(\outprod{00}{11}+ \outprod{11}{00})\nonumber \\&+& w_1
\proj{01}+w_2 \proj{10}+ z \outprod{01}{10}+z^* \outprod{10}{01},
\end{eqnarray}
which is called X states class and arises naturally in a wide
variety of physical situations. If the initial state belongs to
the set of X states (\ref{initial dm}), then Eq. (\ref{spectrum})
guarantees that $\rho(t)$ given by Eq. (\ref{rho t}) also belongs
to the same set. Therefore, the only non-vanishing  components of
the density matrix in  the standard basis are
\begin{eqnarray}\label{DM component}
%\begin{array}{l}
\rho_{11}(t)  & = & \frac{\mu_+}{2 \eta^2}[2 B^2+(J \chi)^2
(1+\Phi(t))] +\frac{\mu_-}{2 \eta^2}[(J \chi)^2
(1-\Phi(t))]-\frac{\nu}{\eta^2}[B J \chi (1-\Phi(t))]\,,
\nonumber\\
\rho_{22}(t)  & = &\frac{w_1}{2 \xi^2}[2 b^2+(J^2+(J_z D)^2)
(1+\Phi'(t))]
+\frac{w_2}{2 \xi^2}[(J^2+(J_z D)^2) (1-\Phi'(t))] \nonumber\\
&-&[\frac{z}{2\xi^2}[(J-iJ_z D)b\, (1-\Psi'(t))]
+C.C]\,, \nonumber\\
\rho_{33} (t) &=& \frac{w_1}{2 \xi^2}[(J^2+(J_z D)^2)
(1-\Phi'(t))]
+\frac{w_2}{2 \xi^2}[2 b^2+(J^2+(J_z D)^2) (1+\Phi'(t))] \nonumber\\
&+& [\frac{z}{2\xi^2}[(J-iJ_z D)b\, (1-\Psi'(t))]
+C.C]\,,\nonumber\\
\rho_{44} (t) &=&  \frac{\mu_+}{2 \eta^2}[(J \chi)^2 (1-\Phi(t))]
+\frac{\mu_-}{2 \eta^2}[2 B^2+(J \chi)^2
(1+\Phi(t))]+\frac{\nu}{\eta^2}
[B J \chi (1-\Phi(t))]\,,\nonumber\\
\rho_{14}(t)   &=& \frac{\mu_+}{2 \eta^2}[BJ\chi(-1+\Psi(t))]
+\frac{\mu_-}{2 \eta^2}[BJ\chi(1-\Psi(t))]+\frac{\nu}{\eta^2}[(J
\chi)^2+B^2 \Psi(t)] \,,
\nonumber\\
\rho_{23}(t)  & =& \frac{w_1}{2 \xi^2}[(J+iJ_z D)b\,
(-1+\Psi'(t))]
+\frac{w_2}{2 \xi^2}[(J+iJ_z D)b\, (1-\Psi'(t))] \nonumber\\
&+&\frac{z}{2\xi^2}[(J^2+(J_z D)^2)+\Theta(t)]
+\frac{z^*}{2\xi^2}[(J+iJ_z D)^2(1-\Phi'(t)))] \,, \nonumber\\
\rho_{41}(t)   &=& \rho_{14}(t)^*\,, \nonumber \\
\rho_{32}(t)   &=& \rho_{23}(t)^*\,.
%\end{array}
\end{eqnarray}
Here, we have defined $\Phi(t):=e^{-2 \eta^2 \gamma \, t}\cos{2
\eta t},\,\, \Psi(t):=[\cos{2 \eta t}-\frac{i \eta}{B}\sin{2 \eta
t}]e^{-2 \eta^2 \gamma \, t},\,\, \Phi'(t):=e^{-2 \xi^2 \gamma \,
t}\cos{2 \xi t},\,\, \Psi'(t):=[\cos{2 \xi t}-\frac{i
\xi}{b}\sin{2 \xi t}]e^{-2 \xi^2 \gamma \, t}$ and
$\Theta(t):=[(\xi^2+b^2)\cos{2 \xi t}-2ib \xi \sin{2 \xi t}]e^{-2
\xi^2 \gamma\, t}$. For asymptotically large times, all of these
functions vanish, and hence the state of the system at
asymptotically large time limit $\rho^\infty$, can be obtained
easily. Knowing the density matrix $\rho(t)$, we can calculate the
entanglement by negativity:
\begin{eqnarray} \label{neg}
N(\rho)=\max\{-2 \min \{\lambda_1, \lambda_2, \lambda_3,
\lambda_4\}, 0 \}
\end{eqnarray}
where
\begin{eqnarray} \label{lambda00}
\lambda_{1,2}&=& \frac{1}{2}((w_1+w_2) \pm
\sqrt{(\rho_{22}(t)-\rho_{33}(t))^2+4|\rho_{14}(t)|^2})\nonumber\\
\lambda_{3,4}&=& \frac{1}{2}((\mu_++\mu_-) \pm
\sqrt{(\rho_{11}(t)-\rho_{44}(t))^2+4|\rho_{23}(t)|^2},
\end{eqnarray}
are eigenvalues of the partially transposed matrix
$\rho^{T_A}(t)$. The negativity is a function of the model
parameters and the initial conditions. Figs. \ref{figure1} and
\ref{figure2} depict the time variation and asymptotical behavior
of the negativity for product and maximally entangled initial
states, respectively. The results show that, the entanglement
reaches a steady state value after some coherent oscillations for
the times greater than $\frac{2}{\gamma}$. The size of this steady
state value, $N^\infty$, depends on the initial conditions and the
parameters of the model. Fig. \ref{figure1} shows that for the
product initial state $\ket{\psi(0)}=\ket{00}$, the function
$N^\infty$ increases with magnetic field for $B\leq 1$ and then
decreases for $B>1$. Hence, for the case of $B\leq 1$ and
$\ket{\psi(0)}=\ket{00}$, we can suppress the effects of dephasing
by increasing $B$. This figure also shows that, for the case of
product initial state $\ket{\psi(0)}=\ket{01}$, $N^\infty$ is a
decreasing function of $D$, for all values of $D$. Fig.
\ref{figure2} shows that, for the case of maximally entangled
initial states $\ket{\psi(0)}=\frac{\ket{00}+\ket{11}}{\sqrt 2}$
and $\ket{\psi(0)}=\frac{\ket{01}+\ket{10}}{\sqrt 2}$, the
function $N^\infty$ decreases with $B$ and $D$, respectively. This
figure also shows that, in the case of $B=0$, the state
$\ket{\psi(0)}=\frac{\ket{00}+\ket{11}}{\sqrt 2}$ has robust
entanglement. Also the state
$\ket{\psi(0)}=\frac{\ket{01}+\ket{10}}{\sqrt 2}$ has robust
entanglement, when $D=b=0$.

\subsection{Quantum Teleportation}
According to the results of Bowen and Bose \cite{BB}, the standard
teleportation protocol ${\cal{T}}_0$, when used with two-qubit
mixed state of the Heisenberg spin chain $\rho(t)$ as a resource,
acts as a generalized depolarizing channel
$\Lambda_{{\cal{T}}_0}^{\rho(t)}[\rho_{in}]$. In the standard
teleportation protocol an input state is destroyed and its replica
(output) state appears  at remote place after applying a local
measurement and unitary transformation in the form of linear
operators. We consider as input an arbitrary pure state
$\ket{\psi_{in}}= \cos\frac{\theta}{2} \ket 0+e^{i \phi} \sin
\frac{\theta}{2} \ket 1$($0\leq \theta \leq \pi$, $0 \leq \phi < 2
\pi$). The output (replica) state, $\rho_{out}$, can be obtained
by applying joint measurement and local unitary transformation on
the input state $\rho_{in}$. Thus the output state is given by
\begin {eqnarray} \label{rho-output1}
\rho_{out}=\Lambda_{{\cal{T}}_0}^{\rho(t)}\,[\,\rho_{in}]=
\sum_{\mu=1}^4 p_\mu\, \sigma^\mu\, \rho_{in}\, \sigma^\mu,
\end {eqnarray}
where $ \mu=0,x,y,z$ ($\sigma^0=I$), $p_{\mu}= Tr [ E^{\mu}
\rho(t)]$ represents the probabilities given by the maximally
entangled fraction of the resource $\rho (t)$. Here
$E^0=\proj{\Psi^-}$, $E^1=\proj{\Phi^-}$, $E^2=\proj{\Phi^+}$ and
$E^3=\proj{\Psi^+}$ where $\ket{\Psi^\pm}=\frac{(\ket{01} \pm
\ket{10})} {\sqrt{2}}$ and $\ket{\Phi^\pm}=\frac{(\ket{00} \pm
\ket{11})} {\sqrt{2}}$ are the Bell states.

The quality of the teleportation is characterized by the concept
of fidelity. The maximal teleportation fidelity achievable in the
standard teleportation protocol ${\cal {T}}_0$ is given by
\cite{HHH}
\begin {eqnarray} \label{max fidelity}
\Phi_{max}(\Lambda_{{\cal{T}}_0}^{\rho(t)})=\frac{2
{\cal{F}}(t)+1}{3},
\end {eqnarray}
where ${\cal{F}}(t)=\max\limits_{\mu=0,1,2,3}\{p_\mu \}$ is the
fully entangled fraction of the resource. For our model, the
probabilities, $p_\mu$s can be written as
\begin{eqnarray} \label{ps}
p_{0, 3}&=&\half (w_1+w_2) \pm \Re e[{\rho_{23}(t)}], \nonumber\\
p_{1, 2}&=&\half(\mu_++\mu_-) \pm \Re e[\rho_{14}(t)].
\end{eqnarray}
Therefore, the maximum fidelity depends on both the initial
conditions of the quantum channel and the parameters of the model.
In the following let's examine some important entangled and
product initial states for the channel:

\emph{i}) $\ket{\psi(0)}_{channel}=\frac{\ket{01} \pm
\ket{10}}{\sqrt 2}$, in this case we have $w_{1,2}=\half$, $z=\pm
\half$ and $\mu_\pm=\nu=0$ and hence
\begin{eqnarray} \label{phi_m1}
\Phi_{max}(\Lambda_{{\cal{T}}_0}^{\rho(t)})=\frac{2}{3}+\frac{1}{3}
\frac{J^2+(b^2+(J_z D)^2)\Phi'(t)}{\xi^2}.
\end{eqnarray} For the
asymptotically large times $\Phi'(t)$ vanishes and we have
\begin{eqnarray} \label{phi_m1-asym.}
\Phi_{max}^\infty=\Phi_{max}(\Lambda_{{\cal{T}}_0}^{\rho^\infty})
=\frac{2}{3}+\frac{1}{3} (\frac{J}{\xi})^2.
\end{eqnarray} This
equation states that, the maximum fidelity achievable at large
time limit is always greater than $2\over 3$ i.e. this channel is
superior to the classical channels. The function $
\Phi_{max}^\infty$ is minimum ($ \Phi_{max}^\infty=\frac{2}{3}$),
if the interaction on the resource is Ising type in the z
direction (i.e. J=0) and reaches its maximum
($\Phi_{max}^\infty=1$) for the case of $D=b=0$. Note that in the
later case the state of the channel is a maximally entangled state
(see Fig. \ref{figure2}).

\emph{ii}) $\ket{\psi(0)}_{channel}=\frac{\ket{00} \pm
\ket{11}}{\sqrt 2}$, i.e. $\mu_\pm=\half$, $\nu=\pm \half$ and
$w_{1,2}=z=0$. The maximum fidelity achievable for this quantum
channel is
\begin{eqnarray} \label {phi_m2}
\Phi_{max}(\Lambda_{{\cal{T}}_0}^{\rho(t)})
=\frac{2}{3}+\frac{1}{3}(\frac{J
\chi}{\eta})^2+\frac{1}{3}(\frac{B}{\eta})^2 \Phi(t),
\end{eqnarray}
and hence,
\begin{eqnarray} \label{phi_m2-asym.}
\Phi_{max}^\infty=\Phi_{max}(\Lambda_{{\cal{T}}_0}^{\rho^\infty})
=\frac{2}{3}+\frac{1}{3}(\frac{J \chi}{\eta})^2,
\end{eqnarray}
which means that, the XY and XYZ chains ($J, \chi \neq 0$) are
more useful resources for performance of the  teleportation
protocol ${\cal T}_0$. In this case, we have $\Phi_{max}^\infty=1$
for $B=0$, this result is compatible with the results of Fig.
(\ref{figure2}).

\emph{iii}) $\ket{\psi(0)}_{channel}=\ket{01}$ (or $\ket{10}$), in
this case just $w_1=1$ ($w_2=1$) is nonzero and hence
\begin{eqnarray} \label{phi_m3}
\Phi_{max}(\Lambda_{{\cal{T}}_0}^{\rho(t)}) =\frac{2}{3}+
|(\frac{b J (1-\Phi'(t))}{3 \xi^2}-\frac{J_z D}{2 \xi}\sin 2\xi t
\,\, e^{-2\xi^2 \gamma t})|.
\end{eqnarray}
At the large time limit we can write
\begin{eqnarray} \label{phi_m3-asym.}
\Phi_{max}^\infty=\Phi_{max}(\Lambda_{{\cal{T}}_0}^{\rho^\infty})
=\frac{2}{3}+\frac{|b J|}{3 \xi^2}.
\end{eqnarray}
and in this case, we can adjust the quality of quantum
teleportation by changing $b$, $J_z$ and $J$. The asymptotic
fidelity, $\Phi_{max}^\infty$, tends to $2 \over 3$ from above for
large values of $b$, thus in the case of $J\neq0$ our channel is
superior to the classical communication. In this case, increasing
$|Jz|$, decreases the quality of teleportation, and hence the XY
chain is more suitable than XYZ chain. There is no way to reach
the value $\Phi_{max}^\infty=1$, since the equation $|b J|=\xi^2$
has no real solution.

\emph{iv}) $\ket{\psi(0)}_{channel}=\ket{00}$ (or $\ket {11}$),
i.e. $\mu_+=1$ ($\mu_-=1$). In this case we have,
\begin{eqnarray} \label{phi_m4}
\Phi_{max}(\Lambda_{{\cal{T}}_0}^{\rho(t)})=\frac{2}{3}+ \frac{B
|J \chi (1-\Phi(t))|}{3 \eta^2},
\end{eqnarray}
and for asymptotically large time we have
\begin{eqnarray} \label{phi_m4-asmp.}
\Phi_{max}^\infty=
\Phi_{max}^\infty(\Lambda_{{\cal{T}}_0}^{\rho^\infty})
=\frac{2}{3}+ \frac{B |J \chi |}{3 \eta^2}.
\end{eqnarray}
According to the Eq. (\ref{phi_m4-asmp.}), the presence of
anisotropy in XY-plane ($\chi\neq0$) provides the desirable
fidelity ($\Phi^\infty >\frac{2}{3}$). The fidelity cannot take
the maximum value $\Phi_{max}^\infty=1$, because the equation $B
|J \chi|=\eta^2$ has no solution in the domain of real numbers.

\begin{figure}
\epsfxsize=16cm \ \centerline{\hspace{0cm}\epsfbox{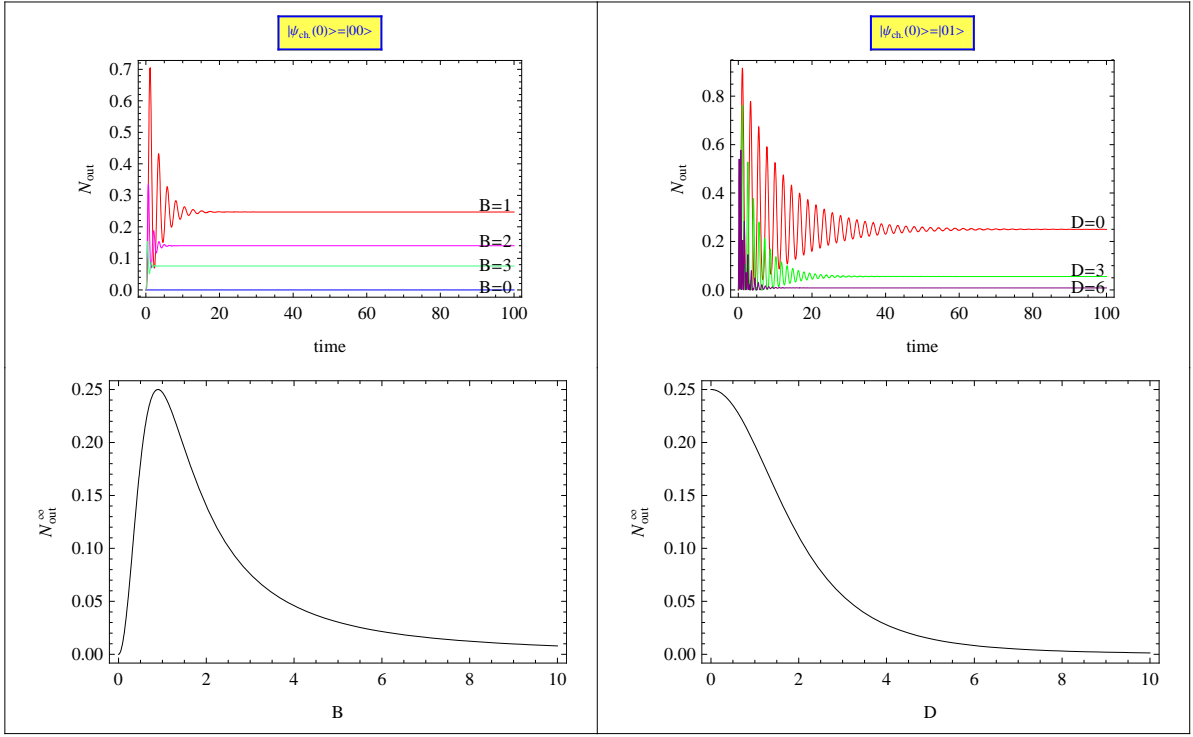}} \
\caption{(Color online) Dynamical and asymptotical behavior of the
output entanglement of the entanglement teleportation protocol
${\cal T}_1$ vs. the model parameters and for the product initial
states, $\ket{\psi(0)}_{channel}=\ket{00}$ (left graph) and
$\ket{\psi(0)}_{channel}=\ket{01}$ (right graph). The input state is
considered a maximally entangled state $N_{in}=1$. The parameters
are the same as in Fig. \ref{figure1}.}\label{figure3}
\end{figure}

\begin{figure}
\epsfxsize=16cm \ \centerline{\hspace{0cm}\epsfbox{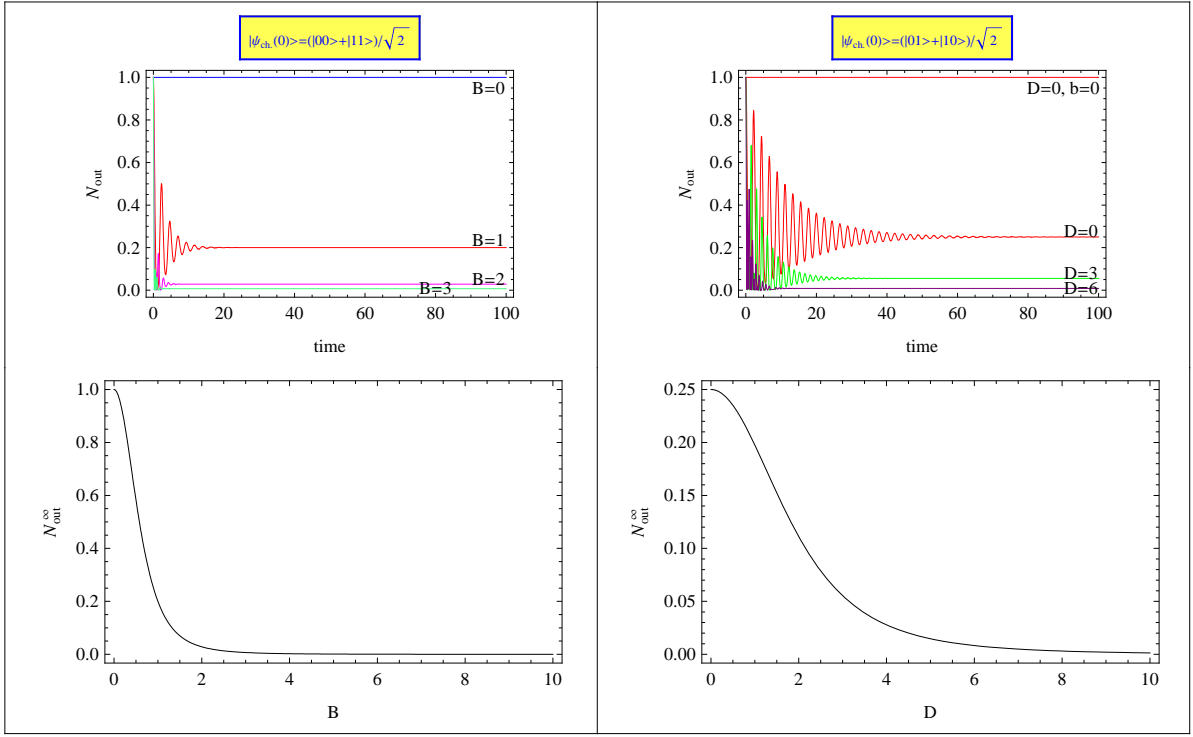}} \
\caption{(Color online) Dynamical and asymptotical behavior of the
output entanglement of the entanglement teleportation protocol
${\cal T}_1$ vs. the model parameters and for the maximally
entangled initial states,$\ket{\psi(0)}_{channel}=\frac{\ket{00} \pm
\ket{11}}{\sqrt 2}$ (left graph) and
$\ket{\psi(0)}_{channel}=\frac{\ket{01} \pm \ket{10}}{\sqrt
 2}$ (right graph). The input state is considered a maximally entangled state $N _{in}=1$.
  The parameters are the same as in
 Fig. \ref{figure1}.}\label{figure4}
\end{figure}

\subsection{Entanglement Teleportation}
\begin{figure}[t]
\epsfxsize=10cm \ \centerline{\hspace{0cm}\epsfbox{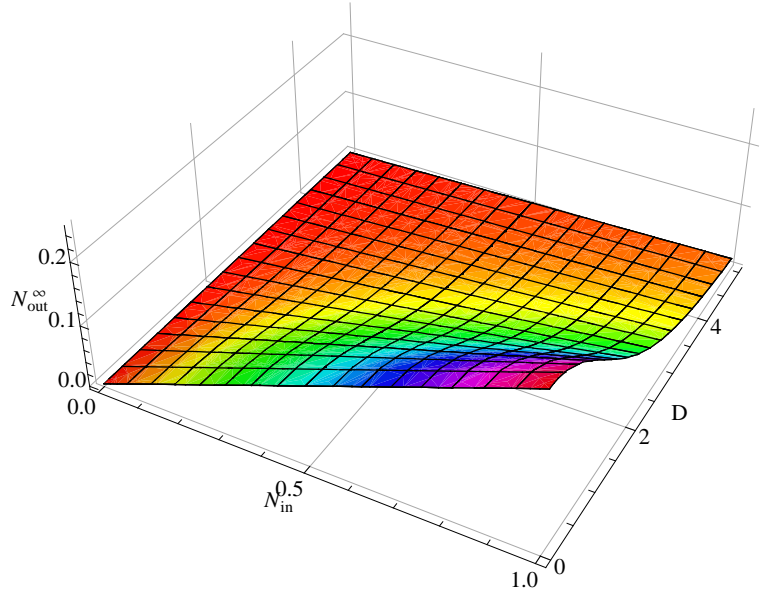}} \
\caption{((Color online) Asymptotic negativity of the output state
of the entanglement teleportation protocol ${\cal T}_1$ vs. $D$ and
the entanglement of the input state, $N_{in}$. The parameters are
the same as in Fig. \ref{figure1}.}\label{figure5}
\end{figure}

In this section, we consider Lee and Kim's teleportation protocol
${\cal{T}}_1$ and use two copies of the above two-qubit state,
$\rho(t)\otimes\rho(t)$, as resource \cite{LK}. We consider the
pure state $\ket{\psi_{in}}=\cos\frac{\theta}{2}\ket{10}+e^{i
\phi} \sin {\theta}{2} \ket{01}$ ($0\leq \theta \leq \pi$, $0 \leq
\phi \leq 2 \pi$) as the input state. The negativity associated
with the input state, $\rho_{in}=\proj{\psi_{in}}$ is
$N(\rho_{in})=N_{in}=\sin \theta$. By generalizing Eq.
(\ref{rho-output1}) the replica (output) state $\rho_{out}$ can be
written as
\begin{eqnarray} \label{output2}
\rho _{out}  =\Lambda_{{\cal T}_1}^{\rho(t)\otimes
\rho(t)}\rho_{in}= \sum\limits_{\mu,\nu} {p_{\mu \nu} (\sigma
_{\mu} \otimes \sigma _{\nu} )} \rho _{in} (\sigma _{\mu} \otimes
\sigma _{\nu}),
\end{eqnarray}
where $\mu=0,x,y,z$, $p_{\mu \nu}=p_\mu p_\nu$. By considering the
two-qubit spin system as a quantum channel, the state of the
channel is given by the equation (\ref{DM component}) and hence
one can obtain $\rho_{out}$ as
\begin{eqnarray}
\rho_{out} &=& \alpha (\proj{00}+\proj{11})+ \kappa(t)
(\outprod{00}{11}+ \outprod{11}{00})\nonumber \\&+& a'
\proj{01}+c'(t) \outprod{01}{10}+c'^*(t) \outprod{10}{01}+b'
\proj{10}
\end{eqnarray}
where
\begin{eqnarray} \label{output2' components}
%\begin{array}{l}
\alpha \,\,\, &=& (w_1  + w_2 )(\mu ^ +   + \mu ^ -  ),\nonumber \\
\kappa(t)  &=& 4\,{\mathop{\rm \Re e}\nolimits} [\rho_{23}(t)]\,
{\mathop{\rm \Re e}\nolimits} [\rho_{14}(t)] \,\cos \phi \,\sin \theta,\nonumber \\
a'\,\,\, &=& (\mu ^ +   + \mu ^ -  )^2 \cos ^2 {\textstyle{\theta
\over 2}}
 + (w_1  + w_2 )^2 \,\sin ^2 {\textstyle{\theta  \over 2}}, \nonumber\\
b'\,\,\, &=& (w_1  + w_2 )^2 \,\cos ^2 {\textstyle{\theta  \over
2}}+ (\mu ^ +   + \mu ^ -  )^2 \sin ^2 {\textstyle{\theta  \over
2}}, \nonumber\\ c'(t) &=& 2\,e^{ - i\phi } (({\mathop{\rm \Re
e}\nolimits} [\rho_{23}(t)])^2 + e^{2i\phi }  ({\mathop{\rm \Re
e}\nolimits} [\rho_{14}(t)])^2 )\,\sin \theta.
%\end{array}
\end{eqnarray}
Now, we can determine the negativity of the output state as
\begin{eqnarray} \label{neg out}
N_{out}=N(\rho_{out})=\max\{-2 \min \{\lambda'_1, \lambda'_2,
\lambda'_3, \lambda'_4 \}, 0 \}
\end{eqnarray}
where,
\begin{eqnarray} \label{lambda'00}
\lambda'_{1,2}&=& \frac{1}{2}((a'+b') \pm
\sqrt{(a'-b')^2+4|\kappa(t)|^2})\nonumber\\
\lambda'_{3,4}&=& \alpha \pm |c'(t)|,
\end{eqnarray}
are the eigenvalues of $\rho_{out}^{T_A}(t)$. The function $N_{out}$
is dependent on the entanglement of the input state $N_{in}$ and the
entanglement of the resource $N_{channel}$ (which is determined by
the initial condition and the parameters of the channel). The
dynamical and asymptotical behavior of the negativity of output
state of the entanglement teleportation protocol ${\cal T}_1$ are
illustrated in Figs. \ref{figure3} and \ref{figure4} for the product
and maximally entangled initial states of the resource,
respectively. In these figures, we assume that the input state is a
maximally entangled state $N_{in}=1$. The results are compatible
with the results of Figs. \ref{figure1} and \ref{figure2}. By
analyzing these figures one can find that, more entangled channel
state is, more input entanglement is preserved. Figure \ref{figure5}
depicts the behavior of the $N_{out}^\infty$ versus the spin-orbit
parameter $D$ (which indicates the entanglement of the channel) and
$N_{in}$, for a given set of other parameters. Increasing the value
of $D$ causes $N_{channel}^\infty$ to decrease (see Figs.
\ref{figure1} and \ref{figure2}) and hence, as the Fig.
\ref{figure5} shows, the entanglement of asymptotic output state,
decreases as $D$ increases, for a fixed value of $N_{in}$. Fig.
\ref{figure5} also shows that, for a fixed value of $D$ (i.e. fixed
$N_{channel}^\infty$), $N_{out}^\infty$ is an increasing function of
$N_{in}$. This means that as the input entanglement increases, a
more entangled quantum channel is required to realize efficient
entanglement teleportation.

The fidelity between $\rho_{in}$ and $\rho_{out}$ in terms of
input negativity ($N_{in}$) is obtained as \cite{RJ}
\begin{eqnarray} \label{fidelity1}
F(\rho _{in} ,\rho _{out};t) = |\bra{\psi_{in}}
\rho_{out}{t}\ket{\psi_{in}}|={\textit{f}}_1(t) +{\textit{f}}
_2(t)\,\, N_{in}^2\,,
\end{eqnarray}
where ${\textit{f}}_1(t)=(w_1+w_2)^2$ and ${\textit{f}}
_2(t)=\frac{1}{2}-(w_1+w_2)+2 ((\Re e[\rho_{14}(t)]) ^2 \cos 2
\phi +(\Re e[\rho_{23}(t)])^2)$. For $\phi=0$, these functions
depend only on the parameters of the channel. This formula has
been also reported in Ref. \cite{LK} by Kim and Lee, but contrary
to the Werner states, ${\textit{f}}_2(t)$ can be a positive number
for Heisenberg chains. This means that, there exists a channel
which teleports more entangled initial states with more fidelity,
but it should be noted that, if we choose the parameters of the
channel such that ${\textit{f}}_2(t)>0$ then ${\textit{f}}_1(t)$
decreases and ultimately $F(\rho _{in} ,\rho _{out};t)$ becomes
smaller than $\frac{2}{3}$, which means that the entanglement
teleportation of mixed states is inferior to classical
communication. Thus, to obtain the same proper fidelity, more
entangled channels are needed for more entangled initial states.

\begin{figure}[t]
\epsfxsize=15cm \ \centerline{\hspace{0cm}\epsfbox{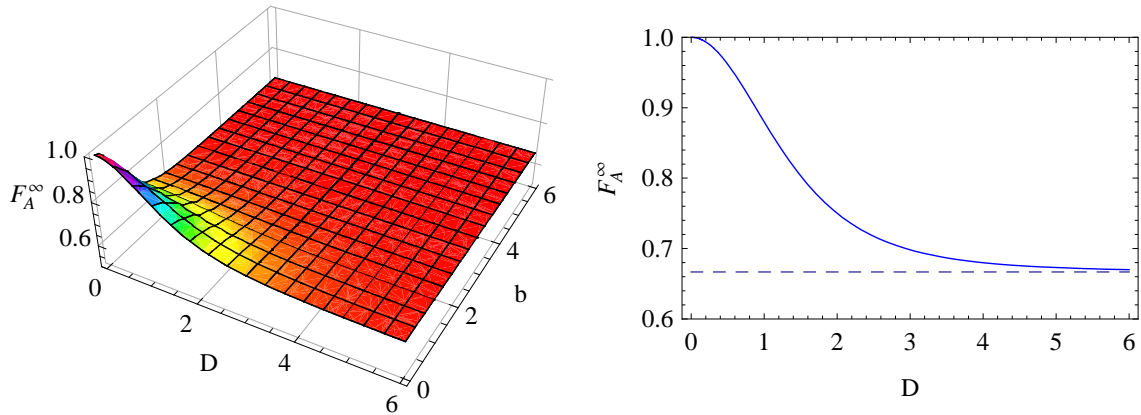}} \
\caption{(Color online) Asymptotic mean fidelity of the entanglement
teleportation protocol ${\cal T}_1$ vs. $D$ and $b$ for maximally
initial state, $\frac{\ket{01} + \ket{10}}{\sqrt 2}$ of the
resource. The parameters are the same as in Fig. \ref{figure1} and
the value of $b=0$ is chosen for the right graph.}\label{figure6}
\end{figure}

 The average fidelity $F_A$ is another useful concept for characterizing the quality of
 teleportation. The average fidelity $F_A$ of teleportation can be obtained by
averaging $F(\rho_{in},\rho_{out};t)$ over all possible initial
states
\begin{eqnarray} \label{FA1}
F_A(t) &=& \frac{{\int_0^{2\pi } {d\phi } \int_0^\pi
{F(\rho_{in},\rho_{out};t)\sin \theta d\theta } }}{{4\pi }}
\nonumber
\\&=&\frac{1}{3}(2 (w_1+w_2)^2+(\mu_++\mu_-)^2+4 (\Re
e[\rho_{23}(t)])^2)\,.
\end{eqnarray}
The function $F_A(t)$ depends on the initial conditions and
parameters of the channel. According to this formula the
asymptotic value of the average fidelity of entanglement
teleportation becomes
$F_A^\infty=\frac{2}{3}(1+\half(\frac{J}{\xi})^4)$, if the channel
is initially in the state, $\ket{\psi(0)}_{channel}=\frac{\ket{01}
+ \ket{10}}{\sqrt 2}$. Fig. \ref{figure6} gives a plot of
$F_A^\infty$ in terms of the parameters  $D$ and $b$ (which
determine $N_{channel}$), in this case. The figure shows that, for
fixed values of the other parameters and $b$, $F_A^\infty$
decreases as $D$ increases (or equally, $N_{channel}$ decreases),
such that for the large values of $D$, $F_A^\infty$ approaches $ 2
\over 3$ from above. We can achieve perfect entanglement
teleportation ($F_A^\infty=1$) in the case of $D=b=0$. For the
case of product initial state, $\ket{\psi(0)}_{channel}=\ket{01}$
of the resource, we have $F_A^\infty=\frac{2}{3}(1+\half(\frac{b
J}{\xi^2})^2)$ which tends to $2 \over 3$ from above for large
values of $D$, too. But never reaches the value 1 for any choice
of the parameters.

\section{DISCUSSION}
The effects of dephasing due to intrinsic decoherence on the
entanglement dynamics of an anisotropic two-qubit Heisenberg XYZ
system in the presence of an inhomogeneous magnetic field and SO
interaction, are investigated. The usefulness of such systems for
performance of the quantum teleportation and entanglement
teleportation protocols are also studied. Intrinsic decoherence
destroys the quantum coherence (and hence quantum entanglement) of
the system as the system evolves. For the case of noninteracting
qubits dephasing processes kill the quantum correlations
(entanglement) of the system at a finite time and hence
entanglement sudden death (ESD) phenomenon occurs (i.e.
entanglement vanishes faster than local coherence of the system
\cite{DT,T}). The results of this paper shows that, for
interacting qubits, dephasing induced by intrinsic decoherence is
competing with inter-qubit interaction terms to create a steady
state level of entanglement after some coherence oscillation and
hence the entanglement of the system reaches a stationary value,
asymptotically. The dynamical and asymptotical behavior of the
entanglement depends on the initial conditions and the system
parameters. Indeed, the effects of dephasing can be amplified or
weakened by adjusting the parameters of the model and initial
conditions. We show that for the product and maximally entangled
initial states, the asymptotic value of the entanglement decreases
as $D$ increases. This is because due to hermiticity of the
Hamiltonian, we can express the state of the system as a
superposition of the energy eigenstates. Increasing $D$, increases
the energy separation of the superposed states (see Eq.
(\ref{eigenvalues})) and hence amplifies the effects of dephasing
(see Eq. (\ref{rho dot 2})). Consequently, for both product and
maximally entangled initial states of the resource, the fidelity
of teleportation approaches $2 \over 3$ form above for large
values of $D$, this is the maximum fidelity for classical
communication of a quantum state. Furthermore, our results show
that, for product initial states and a specific interval of the
magnetic field $B$, the asymptotic entanglement (and hence the
fidelity of the teleportation) can be enhanced by increasing $B$.
We also have argued that, a minimal entanglement of the resource
is required to realize efficient entanglement teleportation. The
results also show that, the XY and XYZ Heisenberg interaction can
provide this  minimal entanglement for the channel state.

We have also found that, the thermal state of the system,
$\rho_T=\frac{e^{-\beta H}}{tr[e^{-\beta H}]}$ is immune to
intrinsic decoherence and hence it has robust entanglement with
respect to the intrinsic decoherence. In the absence of magnetic
field ($B=0$), the maximally entangled initial state,
$\ket{\psi(0)}=\frac{\ket{00} \pm \ket{11}}{\sqrt 2}$, are immune
to intrinsic decoherence and consequently, have robust
entanglement. The same result is also true for the maximally
entangled initial states $\ket{\psi(0)}=\frac{\ket{01} \pm
\ket{10}}{\sqrt 2}$ in the absence of SO interaction ($D=0$) and
for homogeneous magnetic field ($b=0$). Therefore choosing a
proper set of parameters and employing one of these robust states
as initial state of the resource, enable us to perform the quantum
teleportation protocol ${\cal T}_0$ and the entanglement
teleportation ${\cal T}_1$ with perfect quality
($\Phi_{max}=F_A=1$).
\\
\textbf{Acknowledgment}

The authors wish to thank The Office of Graduate Studies and
Research Vice President of The University of Isfahan for their
support.

\newpage
%Figures%%%%%%%%%%%%%%%%%%%%%%%%%%%%%%%%%%%%%%%%%%%%%%%%%%%%%%%%%%%%%%%%%%%%%%%%%%%%%

%end of figures%%%%%%%%%%%%%%%%%%%%%%%%%%%%%%%%%%%%%%%%%%%%%%%%%%%%%%%%%%%%%%%%%%%%%%%%%%%%%%%%%%%%%%%%
\end{document}